\documentclass[twoside,11pt,a4paper]{article}
\usepackage{amssymb}
\usepackage{bm}
\usepackage{amsmath}

\usepackage{color,pstcol}

\begin{document}
\normalsize
\sloppy

\begin{center}
\textbf{Solving the Zeh problem about the density operator with higher-order
statistics}

Alain Deville

\dag Aix-Marseille Universit\'{e}, CNRS, IM2NP UMR 7334, F-13397 Marseille,
France

Yannick Deville

*Universit\'{e} de Toulouse, UPS, CNRS, CNES, OMP, IRAP (Institut de
Recherche en Astrophysique et Plan\'{e}tologie),

F-31400 Toulouse, France

\ \ \ 
\end{center}

\textbf{Abstract \ }

Since a 1932 work from von Neumann, it is generally considered that if two
statistical mixtures are represented by the same density operator $\rho ,$
they should in fact be considered as the same mixture. In a 1970 paper, Zeh,
considering this result to be a consequence of what he called the
measurement axiom, introduced a thought experiment with neutron spins and
showed that in that experiment the density operator could not tell the whole
story. Since then, no consensus has emerged yet, and controversies on the
subject still presently develop. In this paper, stimulated by our previous
works in the field of Quantum Information Processing, we show that the two
mixtures imagined by Zeh, with the same $\rho ,$ should however be
distinguished. We show that this result suppresses a restriction unduly
installed on statistical mixtures, but does not affect the general use of $%
\rho ,$ e.g. in quantum statistical mechanics, and the von Neumann entropy
keeps its own interest and even helps clarifying this confusing consequence
of the measurement axiom. In order to avoid any ambiguity, the
identification of the introduction of this postulate, which von Neumann
rather suggested to be a general property, is given in an appendix where it
is shown that Zeh was right when he spoke of a measurement axiom and
identified his problem. The use and content of a density operator is also
discussed in another physical case which we are led to call the
Landau-Feynman situation, and which implies the concept of entanglement
rather than the one of mixed states.%

\vspace{10mm}
\textbf{Keywords \ }

statistical mixture; density operator; measurement axiom; von Neumann
entropy; improper mixture; higher-order statistics.

\section{Introduction\label{SectionIntroduction}}

The developments of Physics, Communications and Electronics have led to the
birth and growing of a Theory of Information, first in the classical context
(see e.g. the appearance of the Shannon entropy \cite{Shannon1948}) and, for
several decades, in the quantum domain (see e.g. the \textit{Feynman\
Lectures on Computation} \cite{Feynmann1996OnComputation},\textit{\ }and 
\textit{Quantum Computation and Quantum Information }by Nielsen and Chuang 
\cite{NielsenChuang2005}).\ It is now spoken of a second quantum revolution,
which also stimulates a reflection on some basic ideas of Quantum Mechanics
(QM).\ Quantum Information Processing (QIP) is a significant part of the
Quantum Information field. Having been working in the area of QIP for more
than fifteen years (see e.g. \cite{DevilleYAPRA2020}, \cite{DevilleY2023QIP}
and references therein), we were led to introduce the Random-Coefficient
Pure State (RCPS) concept, and not to make an explicit use of the density
operator formalism.\ We however had to think about the content of the
density operator $\rho ,$ and about a question asked by Zeh more than fifty
years ago \cite{Zeh1970}, which we will call the Zeh problem, and which is
discussed in this document.

The present paper uses standard Quantum Mechanics (QM). As a result of its
postulates, including the existence of a principle of superposition (of
states), which the late Nobel Laureate Steven Weinberg called the first
postulate of QM \cite{Weinberg2013}, then, given a quantum system $\Sigma $,
and its state space $\mathcal{E}$, a Hilbert space, any vector of $\mathcal{E%
}$ (defined up to a phase factor $e^{i\varphi }$, $\varphi $ being a real
quantity) represents a possible state of $\sum $ called a pure state. This
standard Hilbert space framework is used by both the so-called orthodox
interpretation of QM (Bohr, Heisenberg, Pauli, Rosenfeld) and by the
statistical interpretation (Einstein, Schr\"{o}dinger, Blokhintsev,
Ballentine), with the meaning given by Ballentine \cite{Ballentine1970} to
the latter expression, one of these interpretations more or less implicitly
accepted by many users of QM. Weinberg has stressed that "quantum field
theory is based on the same quantum mechanics that was invented by Schr\"{o}%
dinger, Heisenberg, Pauli, Born, and others in 1925-26, and has been used
ever since in atomic, molecular, nuclear, and condensed matter physics%
\textit{" }(\cite{WeinbergQuantumFieldsV1}, p.\ 49)\textit{. }We warn the
reader that we are therefore outside the approach initiated by Segal \cite%
{Segal1947}, with his introduction claiming "Hilbert space plays no role in
our theory", an approach known as the C*-algebra formulation of QM, then
developed by Haag and Daniel Kastler, and more recently by Strocchi \cite%
{Strocchi2012} (see also \cite{Drago2018}). We are also outside the approach
from Mielnik \cite{Mielnik} and again Haag, with Bannier \cite%
{HaagNonLinearQM}. The formal constructions and possible results from
mathematical physicists trying to build general quantum theories aiming at
unifying general relativity and Quantum Mechanics (QM), an important field
in present day Physics, are out of the scope of this paper.

A state of the Hilbert space - pure state - used in QM, and described by a
ket in the Dirac formalism \cite{Dirac1939}, obeying the Schr\"{o}dinger
equation if $\Sigma $ is isolated,\ can be obtained from a preparation act,
and if an observable $\widehat{O}$ attached to $\Sigma $ is \textit{measured}
while $\Sigma $ is in the pure (normed) state $\mid \Psi >$, $\ $the mean
value of the result is the quantity $<\Psi \mid $ $\widehat{O}\mid $ $\Psi
>. $\ von Neumann \cite{vonNeumann1932}, \cite{vonNeumann1955English},\ \cite%
{vonNeumann1927} considered a more general situation, called a mixed state
or statistical mixture (of states). Since this von Neumann's work, it is
considered that if two so defined statistical mixtures are represented by
the same density operator they must be seen as the same statistical mixture,
and it is more generally considered that $\rho $ completely describes the
properties of a statistical mixture. Bell's strong reluctance about the
place presently given to measurements in the foundations of QM \cite%
{Bell1989} and his question "Was the wavefunction of the world waiting to
jump for thousands of millions of years until a single-celled living
creature appeared? Or did it have to wait a little longer, for some better
qualified system ... with a PhD?\textit{" \ }are well-known\textit{. }Saying
that his question was provocative is a statement about the question, not an
answer. Already in 1970 Zeh spoke of a \textit{measurement axiom }made by
von Neumann, which led to a \textit{circular argument}, and wrote that: "the
statistical ensemble consisting of equal probabilities of neutrons with spin
up and spin down in the x direction cannot be distinguished by measurement
from the analogous ensemble having the spins parallel or antiparallel to the
y direction. Both ensembles, however, can be easily prepared by appropriate
versions of the Stern-Gerlach experiment. One is justified in describing
both ensembles by the same density matrix as long as the axiom of
measurement is accepted. However, the density matrix formalism cannot be a
complete description of the ensemble, as the ensemble cannot be rederived
from the density matrix\textit{" }\cite{Zeh1970}\textit{.\ }We call this
situation for neutrons proposed by Zeh \textit{the Zeh problem}. Since then,
no consensus has emerged. Recently, for instance, a controversy appeared
after a 2011 paper by Fratini and Hayrapetyan \cite{Fratini2011} claiming
that they had established limits in the statistical operator formalism,
through considerations about variances, followed by a paper from Bodor and
Diosi \cite{Bodor2012}\ asserting that their analysis was irrelevant,
without any final agreement \cite{FratiniArXiv2012}. We recently showed \cite%
{Deville2024} that the use of variances made in \cite{Fratini2011} and \cite%
{FratiniArXiv2012} was wrong.\ The question from Zeh therefore still keeps
its own interest.

One should first clarify the following point: did von Neumann propose a
postulate when introducing the density operator $\rho $, or did he establish
a general property possessed by$\ \rho $? For the sake of brevity, we start
adopting the viewpoint taken by Peres, when, in his book (\cite{Peres1995},
pages 75-76), writing that\textit{\ } "the $\rho $\ matrix completely
specifies all the properties of a quantum ensemble\textit{", }he has spoken
of a\textit{\ "}fundamental postulate\textit{". }We delay a more detailed
analysis of this question towards the interested reader into Appendix 1. In
section \ref{SectionvonNeumannMixture-Postulate}, the introduction of mixed
states and of the density or statistical operator $\rho $ by von\ Neumann is
briefly recalled. In Section \ref{SectionTheZehProblem}, stimulated by our
previous use of higher-order moments in the presence of an RCPS, we come to
the Zeh problem, with a spin 1/2 and \textit{the two von Neumann mixed
states considered by Zeh}, described by the same density operator. Using the
moments of an RV linked to the results of measurements, we show that if the
measured spin component and the RV are both well-chosen, the values of 
\textit{at least one of its moments} differ, when considering these two
mixtures, which allows us to differentiate between these two mixed states.
Section \ref{SectionLandau-Feynman} is devoted to the use of a density
operator in a specific context implying a bipartite quantum system, which we
are led to \ call the Landau-Feynman situation, and\ a concern from
D'Espagnat about the habit of speaking of a mixture in that context is
commented. Section \ref{SectionDiscussion} is devoted to a discussion,
before a conclusion in Section \ref{SectionConclusion}, and two appendices.\
The first one is devoted to what was called the measurement axiom by Zeh.\
In the second one, some information is given about the 1927 paper written by
Landau on the subject presented in Section \ref{SectionLandau-Feynman}.

\section{von Neumann statistical mixture and measurement postulate\label%
{SectionvonNeumannMixture-Postulate}}

We keep the notations introduced in Section \ref{SectionIntroduction}. The
knowledge of a statistical mixture introduced by von Neumann in his 1932
book \cite{vonNeumann1955English} (see also his 1927 paper \cite%
{vonNeumann1927} and Appendix 1) first apppears as the knowledge of a
collection of normed pure states $\mid \varphi _{i}>$ of $\Sigma ,$ each one
with a probability of presence in the mixture equal to $p_{i}$ (for any $i,$%
\ $p_{i}\geq 0$ and $\sum_{i}p_{i}=1$), this collection being collectively
written as \{$\mid \varphi _{i}>$, $\ p_{i}$\}.\ The experiment imagined by
Zeh (see Section \ref{SectionTheZehProblem}) gives an instance of such
mixtures. Note 156 of \cite{vonNeumann1927}, with his reference to von
Mises, indicates that von Neumann adopted what is now called the frequentist
interpretation of probability (see also e.g. \cite{Stacey2016}). In a second
step, $\hat{O}$ being an observable attached to $\Sigma $ and $<\varphi \mid 
\hat{O}\mid \varphi >$ the mean value of $\hat{O}$ in a pure state $\mid
\varphi >,$ von Neumann considered the mean value of $\hat{O}$ in that
statistical mixture, the quantity $\sum_{i}p_{i}<\varphi _{i}\mid \hat{O}%
\mid \varphi _{i}>$. This led him to associate with this statistical mixture
its density operator $\rho =\Sigma _{i}p_{i}\mid \varphi _{i}><\varphi
_{i}\mid ,$ acting linearly on the states\ of $\Sigma $. $\rho $ is
Hermitian, and is positive-definite (all its eigenvalues are non-negative,
see e.g. \cite{Messiah1English}). The eigenvalue spectrum of a Hermitian
positive-definite operator with a finite trace is entirely discrete, a
result of Hilbert space theory (\cite{Messiah1English}, page 335). When an
isolated system is in a statistical mixture, $\rho $ obeys the Liouville-von
Neumann equation. In the special case when $\Sigma $ is in a pure state $%
\mid \Psi >,$ $\rho $ is a projector: $\rho =\mid \Psi ><\Psi \mid .$ The
relation $Tr\rho ^{2}\leq Tr\rho $ is obeyed by $\rho $, the equality being
verified \ iff $\rho $ is a projector, i.e. if and only if $\rho $ describes
a pure state. $\rho ^{2}=\rho $ iff $\rho $ is a projector. von Neumann
started Ch.\ IV of \cite{vonNeumann1955English} saying that in his previous
chapter he had "succeeded in reducing all assertions of quantum mechanics%
\textit{" }to a formula expressing that the mean value of a physical
quantity $O$ when the system is in the state $\mid \Psi >$ is equal to a
quantity written, with our notations, as $<\Psi \mid \widehat{O}\mid \Psi >.$
This passage from \cite{vonNeumann1955English}\ is discussed in Appendix 1.

\section{The Zeh problem and the use of higher-order moments\label%
{SectionTheZehProblem}}

The problem identified by Zeh through his thought experiment manipulating
neutrons was presented in Section \ref{SectionIntroduction}. Zeh introduces
a Stern-Gerlach (SG) equipment.\ In their 1922 experiment, Stern and Gerlach
used silver atoms placed in a furnace heated to a high temperature, leaving
the furnace through a hole and propagating in a straight line. They then
crossed an inhomogeneous magnetic field and condensed on a plate (see \cite%
{Cohen-Tannoudji2020}, page 394). As they have no electric charge, they were
not submitted to the Laplace force, but they have an electronic permanent
magnetic moment. In a classical approach, one should then observe a single
spot, \textit{whereas two spots} were observed, which could only be
explained, later on, as the result of a quantum behaviour: a silver atom has
a spin $1/2.\ $Zeh considers the random emission of neutrons by a neutron
source.\ It is well-established that a neutron has a nuclear spin $1/2$ here
denoted as $\overrightarrow{s}$\ (it is usually written as $\overrightarrow{I%
}$, the symbol $\overrightarrow{s}$ being kept for spins with electronic
origin) and a magnetic moment $\mu =-1.913047$ $\mu _{N}$ ($\mu _{N}:$
nuclear magneton)~proportional to its spin. The force acting on the magnetic
moment of the successive neutrons deflects them into two well-identified
beams, one beam corresponding to the spin quantum state $\mid z,+1/2>$ and
one beam corresponding to the spin quantum state $\ \mid z,-1/2>.$\ The
letter $z$ is reminiscent of the fact that the field gradient and the force
on the spin were directed along $\ z$ in Fig 1, in page 395 of \cite%
{Cohen-Tannoudji2020}. As the neutrons are emitted one by one (no
interaction between them),\ interact only with the magnetic field before
being collected on the plate, and are not each one identified when leaving
the furnace, but are only counted when arriving on the plate, with the same
total number $N/2$ in the two packets, one may say (strictly speaking, in
the limit $N$ $\longrightarrow \infty $) that one prepared the following
(von Neumann) statistical mixture: $\mid +z,1/2>,$ $\frac{1}{2},$ $\mid
-z,1/2>,$ $\frac{1}{2}.\ $This mixture is the one compatible with the SG
equipment in reference \cite{Cohen-Tannoudji2020}.\ Following up the
question from Zeh in \cite{Zeh1970}, we now consider a spin $1/2$,\ and
successively its state in:

\begin{center}
Mixture 1: $\mid +x>,$ $1/2$ and $\mid -x>,$ $1/2,$

Mixture 2: $\mid +y>,$ $1/2$ and $\mid -y>,$ $1/2,$
\end{center}

$\hspace{-0.5cm}\mid +x>$ and $\mid -x>$ being the eigenkets of $s_{x}$ for
the values $+1/2$ and $-1/2$ respectively, and $\mid +y>$ and $\ \mid -y>$
the eigenkets of $s_{y}$ for the values $+1/2$ and $-1/2$ respectively.

The density operator associated with both mixtures is $\rho =I/2$ ($I$:
identity operator in the state space of the spin). We decide to forget the
existence of the von Neumann measurement postulate, which suggests that both
mixtures\ are the same, and which therefore would discourage us from doing
what follows. We choose to use, instead of the $\rho $ formalism, the very
definition of these mixtures. And, in order to try and clarify the Zeh
problem, our previous use of moments in the presence of an RCPS here
suggests us to use moments of an arbitrary order (and not only the mean
value) of a well-chosen RV. Just before the plate, at the level of each
arriving beam, we introduce an equipment able to measure the $\mathbf{s}_{%
\mathbf{x}}$ component of each neutron, and to store the result. Von Neumann
told us that the mean value of the result of this measurement, written in
the Dirac formalism, is:%
\begin{eqnarray*}
\frac{1}{2} &<&+x\mid s_{x}\mid +x>+\frac{1}{2}<-x\mid s_{x}\mid -x>\text{
for mixture 1} \\
\frac{1}{2} &<&+y\mid s_{x}\mid +y>+\frac{1}{2}<-y\mid s_{x}\mid -y>\text{
for mixture 2}
\end{eqnarray*}%
One can interpret these results as the mean value (over all pure states that
compose the considered mixed state) of a Random Variable (RV) which we
denote as $X,$ and which is defined as being itself the \textit{mean }value
taken by $s_{x}$ when the spin is in a given pure state. Its name $X$
recalls us that it mobilizes the $x$ component of the spin. In the specific
case of a pure state $\mid \varphi >,$ $X$ takes the value $<\varphi \mid
s_{x}\mid \varphi >.$

For any value of the non-negative integer $n,$ $\mu _{n},$ the $nth$ moment
\ of $X$ has the following value for mixture 1:%
\begin{equation*}
\text{mixture 1}\text{:}\text{\ }\mu _{n}(X)=\frac{1}{2}(<+x\mid s_{x}\mid
+x>)^{n}+\frac{1}{2}(<-x\mid s_{x}\mid -x>)^{n}
\end{equation*}

\begin{equation*}
\hspace{-3cm}\hspace{1.3cm}=\frac{1}{2}(\frac{1}{2})^{n}+\frac{1}{2}(-\frac{1%
}{2})^{n}
\end{equation*}%
\textit{Therefore, in statistical mixture 1, any odd moment of }$X$ \textit{%
has a value equal to }$0,$\textit{\ and any even moment (}$n$\textit{\ even)
is equal to }$1/2^{n}.$

Considering now mixture $2$, the $nth$ moment \ of $X$ has the value:%
\begin{equation*}
\text{mixture 2: }\mu _{n}(X)=\frac{1}{2}(<+y\mid s_{x}\mid +y>)^{n}+\frac{1%
}{2}(<-y\mid s_{x}\mid -y>)^{n}
\end{equation*}

We recall the developments of $\mid +y>$ and $\mid -y>$ within the standard
basis:%
\begin{equation*}
\mid +y>=\frac{\mid +>+i\mid ->}{\sqrt{2}}\text{ \ \ and \ \ \ }\mid -y>=%
\frac{\mid +>-i\mid ->}{\sqrt{2}}\text{ }
\end{equation*}%
The quantity $<+y\mid s_{x}\mid +y>$ is equal to zero, as the diagonal
quantities $<+\mid s_{x}\mid +>$ and $<-\mid s_{x}\mid ->$ are both equal to 
$0$, and the sum of the interference terms is equal to zero.\ The same
result is obtained for $<-y\mid s_{x}\mid -y>.$

\textit{Therefore, in statistical mixture 2, any moment of }$X$\textit{\ is
equal to 0.}

\textit{We therefore established that all even-order moments of the RV
called X differ when comparing the value they possess in mixture 1 and
mixture 2.}

One guesses that if, in contrast, the same mixtures being considered, one
measures $s_{z}$ instead of $s_{x},$ and one then introduces the RV $Z$,
defined in the same way as $X$ (and which, of course, has nothing to do with 
$Z,$ the direction of a\ magnetic field), the difference found with the
moments of $X$ should disappear with $Z,$ since the choice of $s_{z}$
introduces a new symmetry, and an inability for the new RV $Z$ to
distinguish between the two mixtures through the use of the moments of $%
s_{z} $. We choose to examine this question explicitly. $Z,$ the new RV, is
defined through the way already used for $X,$ $s_{x}$ measurements being
replaced by $s_{z}$ measurements.\ One first considers the values of the
moments of $Z$ when the spin is in mixture $1.$\ The developments of $\mid
+x>$ and $\mid -x>$ in the standard basis are respectively:%
\begin{equation*}
\mid +x>=\frac{\mid +>+\mid ->}{\sqrt{2}}\text{ \ \ and \ \ \ }\mid -x>=%
\frac{\mid +>-\mid ->}{\sqrt{2}}\text{ }
\end{equation*}%
The value of $Z$ in the pure state $\mid +x>$, i.e. $<+x\mid s_{z}\mid +x>,$
when $\mid +x>$ is developed in the standard basis, is obtained as the sum
of its interference terms, each equal to zero, and of the diagonal terms,
the sum of their contributions being equal to $0.\ $Therefore $<+x\mid
s_{z}\mid +x>=0.$\ For the same reason, $<-x\mid s_{z}\mid -x>=0.$
Therefore, any moment of $s_{z}$ in mixture 1 has a value equal to $0.\ $%
Following the same approach, one gets the same result for $Z$ and mixture 2.$%
\ $As expected, considering measurements of $s_{z}$ and the moments of $Z,$
one is unable to establish any difference between Zeh mixtures $1$ and $2.$
This result however does not change the previous conclusion, which
corresponds to a sufficient condition: using two well-chosen mixtures -those
introduced by Zeh- possessing the same density operator, we have been able
to introduce a well-chosen RV, which we called $X,$ related to results of
measurements of a well-chosen spin component, namely $s_{x},$ and such that
at least one of the moments of $X$ had a different value in the two Zeh
mixtures.

\section{The Landau-Feynman situation \label{SectionLandau-Feynman}}

The previous sections were focused on the problem identified by Zeh in the
context of the introduction of the density operator by von Neumann through
the consideration of statistical mixtures.\ In this section, another
physical situation is considered, because it also makes use of a density
operator. We therefore consider a quantum system $\Sigma $, isolated from
its neighbours at the chosen time scale, and composed of $\Sigma _{1},$ the
system of interest, and $\Sigma _{2}.$ At a time $t_{0}$ when $\Sigma _{1}$
and $\Sigma _{2}$ are uncoupled, $\Sigma _{1}$ and $\Sigma _{2}~$are
separately prepared, each in a pure state. $\Sigma $, the global system, is
therefore in a pure state $\mid \Psi (t_{0})>.$\ In a situation when, after
this preparation act, an internal coupling exists between $\Sigma _{1}$\ and 
$\Sigma _{2},$\ and this until some time $t_{1}$,\ one is interested, when $%
t\geq t_{1}$, i.e. once this coupling has disappeared, and at the chosen
time scale, in the mean value of an observable $\widehat{O\text{ }}$ acting
on $\Sigma _{1}$ only. Such a situation was examined by Feynman in Chapter 2
of his \textit{Statistical Mechanics} \cite{Feynman1972}.\ Feynman first
observed that for $t\geq t_{0}$ the whole system obeys the Schr\"{o}dinger
equation. He then showed that this mean value at $t_{1}$ is equal to $%
Tr_{1}\{\rho _{1}(t_{1})\widehat{O}\},$ where $\rho _{1}(t_{1})$ $%
=Tr_{2}\rho (t_{1}),$ $\rho (t_{1})$ being the projector $\mid \Psi
(t_{1})><\Psi (t_{1})\mid $, $\mid \Psi (t_{1})>$ being the ket describing $%
\Sigma $ at $t_{1},$ according to the Schr\"{o}dinger equation, and $Tr_{1}$
being \ a trace calculated over the kets of an orthonormal basis of $\Sigma
_{1}$. He then\ showed that the result keeps true for any time $t>t_{1},$
and finally that the partial trace $\rho _{1}(t)_{\text{ }}$obeys the
Liouville-von Neumann equation for $t\geq t_{1}$. The use of the Schr\"{o}%
dinger equation for $t\geq t_{1}$ for the establishment of this property
implies that when $t\geq t_{1},$ $\Sigma _{1}$ may be submitted to
time-dependent forces giving birth to a time-dependent Hamiltonian, the
sources of these forces (e.g. an oscillating magnetic field acting on a spin
magnetic moment) being then included in $\Sigma _{1}$.

That situation had already been examined by Landau in 1927 \cite{Landau1927}%
, \cite{Landau1927TraduitEnAnglais}, in his own introduction of the density
operator concept. \ Reading the content of \cite{Landau1927} may \ be
difficult for the modern reader, for reasons given in Appendix 2, which
explains our reference to \cite{Feynman1972} and our introduction of the
expression \textit{The Landau-Feynman situation}$.$

Generally, when $t\geq t_{1},$ because of the coupling which did exist
between $\Sigma _{1}$ and $\Sigma _{2}$ until $t_{1},$ the pure state $\mid
\Psi (t)>$ describing $\Sigma $ cannot be written as a product of two kets
describing $\Sigma _{1}$ and $\Sigma _{2}$ respectively. Of course, someone
could take $\Sigma _{2}$ as the system of interest, and introduce his own
so-called reduced statistical operator $\rho _{2}(t).$\ But he is not
allowed to forget that $\Sigma $ is in a pure state $\mid \Psi (t)>,$ and
not allowed to suggest that the state of $\Sigma $ is $\rho _{1}(t)\otimes
\rho _{2}(t),$ generally a mixed state. It is well-known that \textit{the
state of }$\Sigma $\textit{\ is said to be entangled}, a concept first
developed by Schr\"{o}dinger in the context of the 1935\ EPR\ paper.
However, already in a 1939 review paper \cite{LondonBauer1939}%
,
\cite%
{LondonBauerEnglishVersion}, F.\ London and Bauer, examining the same
two-body system, and the same situation, after introducing the two partial
traces, leading to what we have called the density operators $\rho _{1}(t)$
and $\rho _{2}(t),$ wrote: "we see that during the interaction systems I and
II individually transform themselves from pure cases into mixtures.\ This is
a rather strange result".\textit{\ }And even Feynman, in his book \cite%
{Feynman1972}, when treating this question, and considering the case when $%
\rho _{1}^{2}\neq \rho _{1}$ (i.e, the case when the pure state is
entangled) accepted to write (page 41) that the system "is \ in a mixed
state".

A few years before Feynman's book, Bernard d'Espagnat \cite{D'Espagnat1966}
stressed this ambiguity when he made a distinction between a von Neumann
mixture, which he called a \textit{proper mixture} and what we have just
called the Landau-Feynman situation. Speaking of the state of what we denote
as $\Sigma _{1}$ and $\Sigma _{2},$ D' Espagnat wrote that "neither of them
is a pure state\textit{",} and added: "One then usually says that they are a
mixture\textit{" }(see e.g. our reference \cite{LondonBauerEnglishVersion}),
and he then proposed to say instead that one has an \textit{improper mixture.%
} In the following decades, D'Espagnat kept insisting on his distinction
(see e.g.\ his 2001 reply \cite{D'Espagnat2001} to a paper by K.A.
Kirkpatrick \cite{Kirkpatrick2001} and the references in both papers).\ It
is clear that one has to speak of a pure entangled state of $\Sigma _{1}$
and $\Sigma _{2}$, and accepting to speak of a mixture, even while adding
the adjective \textit{improper}, is however making a concession which keeps
an ambiguity.

\section{Discussion \label{SectionDiscussion}}

In a document published in 2004 in Physics Today \cite{Mermin2004}, Mermin
discussed the origin of the expression \textit{Shut up and calculate.}$\ $%
Such an expression aims at describing an attitude possibly adopted when
facing, in QM, the existence of the superposition principle, the content of
the pure state concept, and the related interference phenomenon. None of
them is the subject of this paper. The subject of the present document, far
more modest, is still in debate, and e.g. a 2022 Arxiv paper \cite%
{Castellani2022} is entitled \textit{All quantum mixtures are proper.} That
2022 paper expresses a disagreement with the 2001 article from D'Espagnat,
but starts writing "We also assume that the state of a system is completely
described by its density matrix" and refers to \cite{vonNeumann1955English},
which anyway did not discuss the Landau-Feynman situation. As we cite it, we
should at once add, first, that the approach taken in \cite{Castellani2022}
minimizes the existence of what we called the Landau-Feynman situation and,
secondly, that such an assumption both ignores the existence of the 1970\
paper by Zeh and eliminates any possibility of debating over the subject.

In the Landau-Feynman situation, it is clear that the considered state of
the bipartite system is a pure but entangled state, and that the $\rho $
operator is a tool for \textit{calculating mean values} of observables
attached to the system of interest, in a specific situation which does not
require any interpretation in terms of possible states attributed to the
system of interest.

The von Neumann approach started by assuming the existence of mixed states,
defined using a classical (i.e non quantum) definition of the probability
concept, through its so-called frequency interpretation (cf.\ Appendix 1).
von Neumann decided to calculate the mean value of an observable attached to
the system when the state of this system is described by the mixed state he
had imagined. This led him to introduce a density operator $\rho $.\ Then,
instead of considering that $\rho $ is a tool useful for the calculation of
a mean value, truly an answer to an important question, he tried to do
something diffferent, i.e. to interpret $\rho $ as describing the state of
the system, therefore giving up his first definition of the situation
through his original definition of a mixed state. In Appendix 1, we refer to
the content of \cite{vonNeumann1955English}, which shows that von Neumann
did introduce a postulate when claiming that, in 
the
presence of a mixed state
as he had defined it, one had finally to redefine the state of the system
through its associated density operator $\rho .$

In his 1970 paper focused on the spin of neutrons and Stern-Gerlach
equipments, and on two statistical mixtures chosen so that both mixtures
have the same 
density
operator $\rho =I/2$\ , Zeh observed that the
description with $\rho $ should not tell the whole story for these mixtures,
since it forgets the initial preparation process of these mixtures.\ We have
just: 1) decided to ignore the von Neumann measurement postulate (cf.\
Section \ref{SectionvonNeumannMixture-Postulate} and Appendix 1), 2)
introduced a well-chosen spin-operator, $s_{x},$ and an RV denoted as $X$
and compatible with what one usually says in QM about the mean value of an
observable in the presence of a statistical mixture, 3) established that the
even moments of X have different values in mixture 1 and in mixture 2.\ This
result allows us to say that, contrary to what is claimed when accepting the
von Neumann measurement postulate, these two mixtures should be
distinguished.\ This result is a sufficient property: when two mixtures have
the same density matrix, once the von Neumann postulate has been given up,
one should consider the very definition of a given statistical mixture (i.e.
the existence of the corresponding collection of pure states, each one with
its given probability), and use e.g. a well-defined RV linked to this
mixture, and its moments.\ The associated density operator, certainly an
important tool, does not necessarily contain the whole information contained
in the mixture \{$\mid \varphi _{i}>$, $p_{i}$\}, which confirms an
intuition from Zeh.

In our discussion of the Zeh problem, we did use the fact that, when $\sum $
is in the pure (normed) state $\mid \Psi >,$ the mean value of an observable 
$\widehat{O}$ is the quantity $<\Psi \mid $ $\widehat{O}\mid $ $\Psi >.$
What we did not use is the von Neumann measurement postulate itself (cf.
also Appendix 1), and its consequence that $\rho $ should contain all the
information contained in the definition of a statistical mixture as the \{$%
\mid \varphi _{i}>$, $p_{i}$\} collection. This giving up of the measurement
axiom does not affect the use of $\rho $ and its importance e.g. in quantum
statistical mechanics. von Neumann proposed the quantity $S=-k_{B}<Ln\rho
>=-k_{B}Tr\{\rho Ln\rho \}$ ($k_{B}:$ Boltzmann constant) as a definition of
the entropy, and similarly this is not affected by the giving up of the
measurement axiom. And the use of the von Neumann entropy moreover helps us
in identifying a consequence of the introduction of the measurement axiom.
In \cite{vonNeumann1955English}, when examining the question of the quantum
analog of the classical entropy, von\ Neumann first established that all
pure states have the same entropy, which he took as the origin of entropy.
In an interpretation of the entropy as a measure of disorder this entropic
behavior is understood as the fact that all pure states present the same
quantity of disorder. We have shown that the two neutron mixtures introduced
by Zeh should be distinguished and, since they do possess the same density
operator $\rho ,$ they have the same value of their entropy, and therefore
the same degree of disorder.\ Introducing the axiom of measurement and
therefore claiming that they are the same mixture introduces a confusion
between degree of disorder and true existence.

\section{Conclusion\label{SectionConclusion}}

Our previous works in the field of Quantum Information Processing, and more
especially the development of the Random-Coefficient Pure State (RCPS)
concept%
, 
within standard Quantum Mechanics (QM), have led us to examine
both the links between RCPS and the statistical operator $\rho ,$ to be
treated in a separate paper, and the question of what, in this paper, is
called the Zeh problem. This problem arises as a consequence of what Zeh
called the measurement axiom, introduced by von Neumann. This axiom says
that, in the presence of a statistical mixture \{$\mid \varphi _{i}>,p_{i}$%
\}, because the mean value of an observable $\widehat{O}$ is $%
\sum_{i}p_{i}<\varphi _{i}\mid \hat{O}\mid \varphi _{i}>,$ the whole
information contained in the mixture is also contained in its associated
density operator \ $\rho =\sum_{i}p_{i}\mid \varphi _{i}><\varphi _{i}\mid $%
. Focusing on the thought experiment proposed by Zeh, with neutron spins,
and on two statistical mixtures also introduced by Zeh and possessing the
same density operator, we succeeded in introducing a Random Variable\ (RV)
with the significant property that, whereas its first moment (mathematical
expectation, or mean value) does possess the same value for both statistical
mixtures, at least one higher-order moment, and even all the even-order
moments of that RV have different values for these mixtures, which clearly
means that these two mixtures should be distinguished. The familiar use of $%
\rho $, e.g. in quantum statistical mechanics, or in the definition of the
von Neumann entropy, is not affected by this inadequate introduction of this
measurement axiom. \ Just in contrast, the use of the entropy concept helps
interpreting the confusion resulting from this axiom. To\ put it another
way, our result is not a restriction about the use of $\rho \,,$ but a tool
for avoiding a possible loss of information when manipulating statistical
mixtures. Because of von Neumann's authority, and the not unfrequent opinion
that von Neumann did establish the property that the density operator keeps
the whole story about two mixtures with the same $\rho ,$ and more generally
about statistical mixtures, we introduced an Appendix focused on passages
from the canonical book by von Neumann devoted to this question, which
confirms that Zeh was right when speaking of the measurement axiom. 
The present
paper also discussed what we called the Landau-Feynman situation, which also
introduces a density operator, again in order to calculate a mean value of
an observable in a specific context, and, in the continuation of a concern
previously expressed by Bernard D'Espagnat, we stressed that the right
concept to be used in that situation is the one of entanglement, and surely
not the one of any mixed state.

\section*{%
Appendix 1: the 
von Neumann measurement postulate \label%
{SectionAppendixvonNeumannPostulate}}

Before the discussion of this question, a comment is made about the
definition of a mixed state or statistical mixture. In section \ref%
{SectionvonNeumannMixture-Postulate}, it was stressed that, in his Note 156
from \cite{vonNeumann1955English}, von Neumann indicates that he uses what
is now called the frequentist interpretation of probability. In order to
avoid any misunderstanding for a reader unfamiliar with the mixed state
concept, we first recall the definition given by Cohen-Tannoudji \textit{et
al}\ in \cite{Cohen-Tannoudji2020} (page 300): "\textit{the state of this
system may be either the state }$\mid \psi 
_{1}
>
$ \textit{with a probability }$%
p_{1}$, \textit{or the state }$\mid \psi _{2}>$\textit{\ with a probability }%
$p_{2},$ \textit{etc... Obviously: }$p_{1}+p_{2}+...=$ $\sum_{k}p_{k}=1."$

The existence of von Neumann's measurement postulate was claimed in Section\ %
\ref{SectionIntroduction}, through a citation from Peres, but one still has
to examine this point in more detail and then to identify the reason of its
introduction. When examining von Neumann's conception of a statistical
mixture from \cite{vonNeumann1955English}, a first difficulty is the fact
that this question occupies parts of four of its six chapters.\ A possible
second one, for the modern reader, results from the fact that von Neumann
used the language of the wave function, and obviously not the ket formalism,
introduced by Dirac seven years later \cite{Dirac1939}, and in this Appendix
we consequently both respect his own writing and, when commenting passages
from \cite{vonNeumann1955English}, keep the notations introduced in Section %
\ref{SectionIntroduction}. In \cite{vonNeumann1955English}, von Neumann,
having considered the probability content attached to a pure state, adds
(pages 295-296)\ "the statistical character may become even more prominent,
if we do not even know what state is actually present \ - - for example when
several states $\phi _{1},$ \ $\phi _{2},...$\ with the respective
probabilities $w_{1},$\ $w_{2},...$($w_{1}\geq 0,$ $w_{2}\geq 0,..$.$%
w_{1}+w_{2}+...=1$) constitute the description" of $S$, the quantum system
of interest. He moreover considers\ (page 298)\textit{\ }"great statistical
ensembles which consist of many systems $S_{1}$\ , . . ., $S_{N},$ i.e., N
models of S , N large". Similarly, at the beginning of his Chapter V,
devoted to thermodynamical questions, von Neumann, extending Gibbs' replica
method into the quantum domain, introduces a mental ensemble of identical
systems in which he measures some operator $R$, now separating this ensemble
into sub-ensembles according to the result of the measurement. He has
started Ch.\ IV of \cite{vonNeumann1955English} saying that in his previous
chapter he has "succeeded in reducing all assertions of quantum mechanics"%
\textit{\ }to a formula expressing that the mean value of a physical
quantity $O$ when the system is in state $\mid \Psi >$ is equal to a
quantity written, with our notations, as $<\Psi \mid \widehat{O}\mid \Psi >.$
But this he postulated in his Ch.\ III$,$ as the reader may convince
himself: he has first to see the existence of property $E_{2}$ in page 203
of \ \cite{vonNeumann1955English}, and then, in its page 210, to read that:
"we recognize $P$. (or $E_{2}$\ .) as the most far reaching pronouncement on
elementary processes". But von Neumann has first written: "We shall now
assume this statement P to be generally valid" (page 201), and "We shall now
deduce $E_{1}$. from $P$., and $E_{2}$. from $E_{1}$." (page 203).
Consequently, given a system $\Sigma $ in a statistical mixture described by 
$\rho $ (which von Neumann named as $U$),\textit{\ } and $\widehat{O}$
attached to an observable $O$ of $\sum ,$ the assertion that everything
should be contained in the expression $E\{\widehat{O}\}=Tr\{\rho \widehat{O}%
\}$ expresses a postulate. However, this fact is not always identified, a
result of von\ Neumann's authority.

We still have to try and identify which reason led von Neumann to introduce
that postulate. In the preface of his 1932 book \cite{vonNeumann1955English}%
, von Neumann wrote that, at the time of its writing,\textit{\ }"the
relation of quantum mechanics to statistics and to the classical statistical
mechanics"\textit{\ }was\textit{\ }"of special importance". And 25 years
later Fano \cite{Fano1957} noted that, in that time interval, "States with
less than maximum information, represented by density matrices $\rho $, have
been considered primarily in statistical mechanics and their discussion has
been influenced by the historical background in this field". In the previous
development of classical statistical mechanics, Gibbs had introduced a
probability density (within the phase space), used for the calculations of
mean values. In contrast, what corresponds to what is now called
higher-order moments (see e.g. their use in \cite{DevilleY2023QIP}) had not
been explicitly considered in physics.\ Therefore, when von Neumann
introduced his measurement postulate, this he could implicitly consider not
to be responsible for a loss of information as compared with that contained
in the definition of a statistical mixture through the explicit
consideration of the \{$\mid \varphi _{i}>$, $p_{i}$\} collection.\ And,
more importantly in the context of a building of quantum statistical
mechanics, the indiscernability of identical particles had already been
identified, which led to distinguish between their mathematical Hilbert
space, and a subspace built from either symmetrical states in the exchange
of two particles (bosons) or antisymmetrical states (fermions), and moreover
to build the quantum Maxwell-Boltzmann statistics for independent but
distinguishable particles, e.g. electron spins diluted in ionic solids (cf.
more generally Chapter X of \cite{Tolman1938}).

\section*{%
Appendix 2: on the Landau approach\label{SectionAppendixLandau}}

In his 1927 paper on the subject, Landau (born in 1908), in the first
section of \cite{Landau1927}, entitled "Coupled systems in wave mechanics"%
\textit{, }wrote: "A system cannot be uniquely defined in wave mechanics; we
always have a probability ensemble (statistical treatment). If the system is
coupled with another, there is a double uncertainty in its behaviour\textit{%
". }But an \ operator then introduced through a \textit{Partial Trace}
procedure \textit{in the presence of such a coupling }does not obey the
Liouville-von Neumann equation, and calling it a density operator nowadays
introduces a confusion.\ Landau and Lifshitz, in \cite{Landau1958} (Volume
III of their Course), first supposed that a\textit{\ "}closed system as a
whole is in some state described by a wave function $\Psi (q,x),$\ where $x$
denotes the set of coordinates of the system considered, and $q$ the
remaining coordinates of the system considered\textit{". }Integrating over
the $q$ variables -which corresponds to introducing a partial trace -, they
introduced an operator \textit{which they again called a density matrix}
(thus keeping the difference with its today well-accepted\ meaning resulting
from the von Neumann approach). Then, in a second step only, they "suppose
that the system\textit{" }(of interest) \textit{"}is closed, or became so at
some time\textit{". }As detailed in Section \ref{SectionLandau-Feynman}, in
Chapter 2 of his \textit{Statistical Mechanics} \cite{Feynman1972}, Feynman
suppressed the possible confusion resulting from the use of the expression
density or statistical operator by both von Neumann and Landau under
different assumptions (the possible existence of a coupling of the system of
interest with a second system in Landau's approach)\textit{.}

\section*{Statements and declarations}

The authors declare to have no financial or non-financial conflict of
interest.

\end{document}